\documentclass[journal=ancac3,manuscript=article]{achemso}

\usepackage[version=3]{mhchem} 
\usepackage{upgreek}
\usepackage{gensymb}
\usepackage{xcolor}
\usepackage{multirow}
\usepackage{subfiles}
\usepackage{chapterbib}
\usepackage{chemformula}
\usepackage{graphicx}

\def\equationautorefname~#1\null{Eq.~(#1)\null}

\author{Jonas Berzin\v{s}}
\affiliation[FSU]
{Institute of Applied Physics, Abbe Center of Photonics, Friedrich Schiller University Jena, Albert-Einstein-Str.~15, 07745 Jena, Germany}
\alsoaffiliation[TNO]
{Optics Department, Netherlands Organization for Applied Scientific Research (TNO), Stieltjesweg~1, 2628CK Delft, The Netherlands}
\email{jonas.berzins@uni-jena.de}
\author{Simonas Indri\v{s}i\={u}nas}
\affiliation[FTMC]
{Department of Laser Technologies, Center for Physical Sciences and Technology, Savanoriu Ave.~231, LT-02300 Vilnius, Lithuania}
\author{Koen van Erve}
\affiliation[TNO]
{Optics Department, Netherlands Organization for Applied Scientific Research (TNO), Stieltjesweg~1, 2628CK Delft, The Netherlands}
\alsoaffiliation[TUe2]
{Department of Applied Physics, Eindhoven University of Technology, 5600~MB Eindhoven, The Netherlands}
\author{Arvind Nagarajan}
\affiliation[TNO]
{Optics Department, Netherlands Organization for Applied Scientific Research (TNO), Stieltjesweg~1, 2628CK Delft, The Netherlands}
\alsoaffiliation[TUe1]
{Department of Electrical Engineering, Eindhoven University of Technology, 5600~MB Eindhoven, The Netherlands}
\author{Stefan Fasold}
\affiliation[FSU]
{Institute of Applied Physics, Abbe Center of Photonics, Friedrich Schiller University Jena, Albert-Einstein-Str.~15, 07745 Jena, Germany}
\author{Michael Steinert}
\affiliation[FSU]
{Institute of Applied Physics, Abbe Center of Photonics, Friedrich Schiller University Jena, Albert-Einstein-Str.~15, 07745 Jena, Germany}
\author{Giampiero Gerini}
\affiliation[TNO]
{Optics Department, Netherlands Organization for Applied Scientific Research (TNO), Stieltjesweg~1, 2628CK Delft, The Netherlands}
\alsoaffiliation[TUe1]
{Department of Electrical Engineering, Eindhoven University of Technology, 5600~MB Eindhoven, The Netherlands}
\author{Paulius Ge\v{c}ys}
\affiliation[FTMC]
{Department of Laser Technologies, Center for Physical Sciences and Technology, Savanoriu Ave.~231, LT-02300 Vilnius, Lithuania}
\author{Thomas Pertsch}
\affiliation[FSU]
{Institute of Applied Physics, Abbe Center of Photonics, Friedrich Schiller University Jena, Albert-Einstein-Str.~15, 07745 Jena, Germany}
\alsoaffiliation[FSU]
{Fraunhofer Institute for Applied Optics and Precision Engineering, Albert-Einstein-Str.~7, 07745 Jena, Germany}
\author{Stefan M. B. B\"{a}umer}
\affiliation[TNO]
{Optics Department, Netherlands Organization for Applied Scientific Research (TNO), Stieltjesweg~1, 2628CK Delft, The Netherlands}
\author{Frank Setzpfandt}
\affiliation[FSU]
{Institute of Applied Physics, Abbe Center of Photonics, Friedrich Schiller University Jena, Albert-Einstein-Str.~15, 07745 Jena, Germany}

\title {Direct and High-Throughput Fabrication of Mie-Resonant Metasurfaces via Single-Pulse Laser Interference}


\begin{document}


\newpage
\begin{abstract}
High-index dielectric metasurfaces featuring Mie-type electric and magnetic resonances have been of a great interest in a variety of applications such as imaging, sensing, photovoltaics and others, which led to the necessity of an efficient large-scale fabrication technique. To address this, here we demonstrate the use of single-pulse laser interference for direct patterning of an amorphous silicon film into an array of Mie resonators. The proposed technique is based on laser-interference-induced dewetting. A precise control of the laser pulse energy enables the fabrication of ordered dielectric metasurfaces in areas spanning tens of micrometers and consisting of thousands of hemispherical nanoparticles with a single laser shot. The fabricated nanoparticles exhibit a wavelength-dependent optical response with a strong electric dipole signature. Variation of the pre-deposited silicon film thickness allows tailoring of the resonances in the targeted visible and infrared spectral ranges. Such direct and high-throughput fabrication paves the way towards a simple realization of spatially invariant metasurface-based devices.
\end{abstract}

\newpage
\section{Introduction}
High-index materials, such as silicon (Si), have been established as an alternative to plasmonic materials in the field of nanophotonics~\cite{genevet2017recent, staude2017inspired}. The use of dielectric metasurfaces, two-dimensional arrays of resonant nanoparticles based on low-loss and high-index dielectric materials, provides a direct path to efficient photonic devices for imaging~\cite{berzins2019submicrometer, koirala2017all, horie2017visible, jin2019dielectric}, sensing~\cite{cernigoj2018lattice, caldarola2015nonplasmonic, bosio2019plasmonic}, light emission~\cite{vaskin2019light}, light harvesting~\cite{slivina2019insights}, and many other applications. The elements of dielectric metasurfaces, so-called Mie resonators, support electric and magnetic resonances~\cite{mie1908}, which enable a wavelength-dependent optical response tailored for a specific functionality via control of the elements shape and size~\cite{groep2013designing, staude2013tailoring}. However, up to now, there was no large-area and low-cost fabrication technique for Mie-resonant metasurfaces.

Even-though Si is a well-known and CMOS compatible material, Si metasurfaces are usually fabricated by complex lithographic techniques, such as electron beam lithography followed by reactive ion etching~\cite{berzins2019submicrometer, arslan2017angle} or focused ion beam lithography~\cite{erdmanis2014focused}. Despite the precision offered, these techniques are limited to a small scale due to low-throughput and high cost of the process. Other pattern transfer techniques like interference lithography~\cite{lee2006fabrication, debor2010sub}, mask aligner lithography~\cite{vetter2018printing}, nanosphere lithography~\cite{lu2003nanopatterning}, or nanoimprint~\cite{chou2002ultrafast} offer a greater flexibility in scaling-up but require multiple lengthy steps. In contrast, chemical synthesis~\cite{shi2013monodisperse} can be a relatively fast and large-scale technique but cannot ensure the periodic distribution of nanoparticles, similar to laser ablation in liquids~\cite{simakin2004nanoparticles}, grinding and milling~\cite{chaabani2019large}, or spontaneous dewetting~\cite{abbarchi2014wafer, thompson2012solid}. An alternative approach was introduced by using pulsed laser radiation, which by means of point-by-point material transfer~\cite{zywietz2014laser, makarov2018resonant} or direct writing~\cite{dmitriev2016laser} enables the formation of ordered metasurfaces. These methods are direct and provide control of the periodicity, but their efficiency is limited to the repetition rate of the pulsed laser, as only a single metasurface element is produced by each laser pulse. To overcome this limitation and obtain high-throughput and cost-effectiveness, we propose the application of direct laser interference patterning.
	
Direct patterning using high-peak-power laser interference has been previously shown as a suitable tool for large-area microstructuring of metals~\cite{alessanria2008direct, voisiat2011flexible, nakata2012interfering, nakata2015fabrication, indrisiunas2015direct, bieda2016fabrication, nakata2018local, voisiat2019improving, indrisiunas2013two, aguilar2018micro}, organic materials~\cite{zhai2011direct,zhai2015direct, stankevicius2019mechanism},
and even Si~\cite{lorens2012micropatterning, vinciunas2013effect, wang2013direct, oliveira2013sub, hu2016bio, indrisiunas2017new, indrisiunas2014influence, gedvilas2018nanoscale}. It does not require resist, etching or any other post-processing steps, thus is relatively fast and simple. Although direct laser interference patterning has been used to achieve submicrometer resolution, as shown in the case of a bulk Si~\cite{oliveira2013sub, gedvilas2018nanoscale}, the technique has never been exploited to obtain high-index metasurfaces.
	
In this paper, laser interference is applied for direct patterning of Si films. It is demonstrated that Si-based Mie-resonant metasurfaces can be fabricated in a single step by single-pulse laser interference. The obtained submicrometer Mie resonators are tailored via control of irradiation conditions and initial film thickness. The structural and spectral analysis of the fabricated metasurfaces is used for the evaluation of scalability and applicability of the proposed technique. We show that single-pulse laser interference is a technique for a direct and high-throughput fabrication of dielectric metasurfaces, and we introduce guidelines for its potential applications.

\section{Results and discussion}
\subsection{Concept}
The mechanism of laser interference patterning of thin films can be understood as a templated dewetting, but instead of pre-patterning by ion milling~\cite{lian2006patterning, naffouti2016templated}, electron beam lithography~\cite{fowlkes2011self}, or photolithography and subsequent wet etching~\cite{ye2011templated}, the interference pattern itself is used as a template for the required configuration of nanostructures.

In general, interference of electromagnetic waves appears when two or more coherent laser beams overlap with each other. For a number of beams $N$ with the same optical frequency, the interference intensity profile can be expressed as follows~\cite{hecht2001optics}:
\begin{equation}
I(\mathbf{r}) 
\propto \frac{1}{2}\sum\limits_{i=1}^N\vert \mathbf{E}_{0i} \vert^2 + \sum\limits_{j<i}^N\sum\limits_{i=1}^N \mathbf{E}_{0i} \cdot \mathbf{E}_{0j} \times \cos(\mathbf{k}_{i} \cdot \mathbf{r} - \mathbf{k}_{j} \cdot \mathbf{r} + \varphi_i - \varphi_j)~,
\end{equation} 
where $\mathbf{E}_{0i}$ and $\mathbf{E}_{0j}$ represent the electric fields, $\mathbf{k}_{i}$ and $\mathbf{k}_{j}$ are the wavevectors, and $\varphi_i$ and $\varphi_j$ are the phases of the respective beams. 

\begin{figure}[!h]
\centering
\includegraphics[width=165mm]{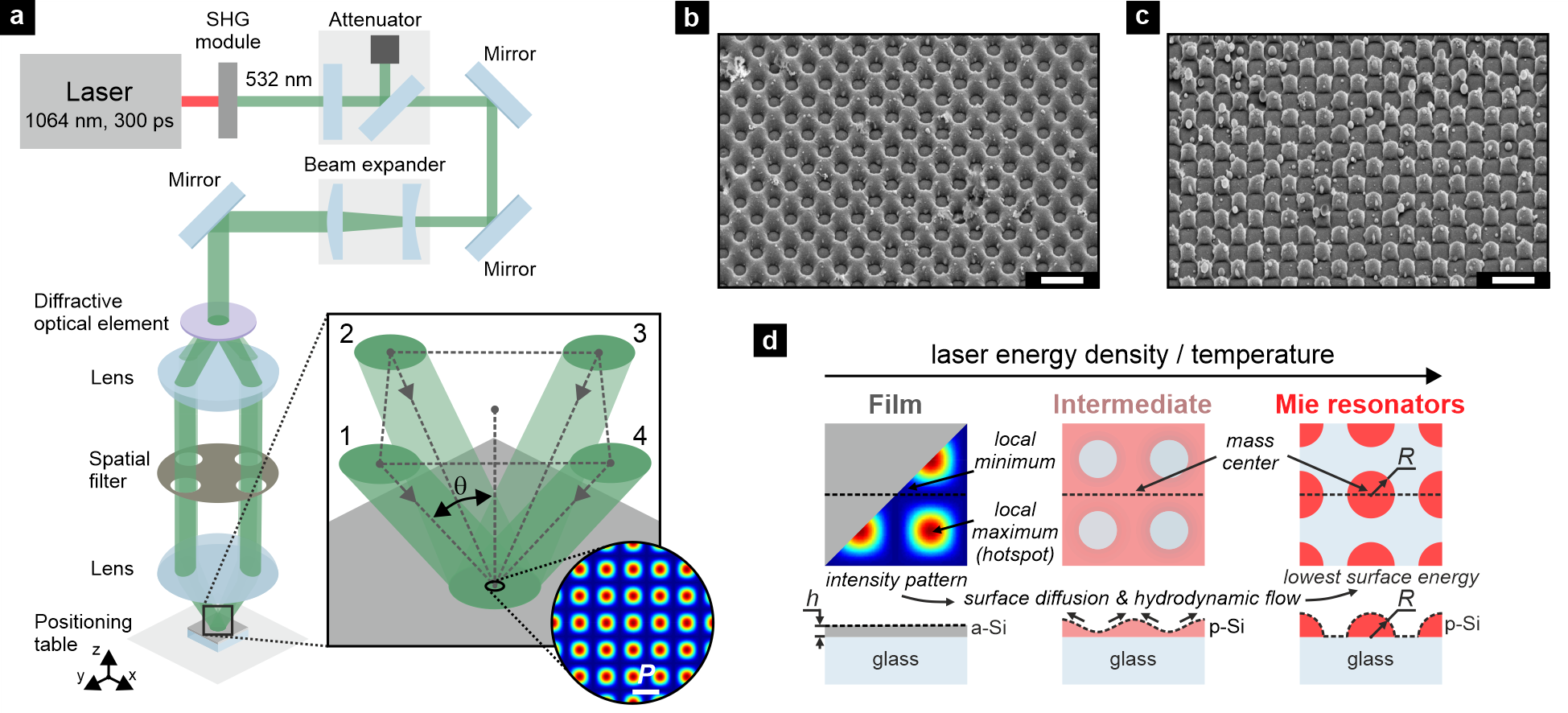}
\caption{Concept of patterning Mie resonant metasurfaces by single-pulse laser interference. (a)~Schematic representation of the four-beam interference setup. The diffractive element splits the beam into four, which are further propagated through a lens system to the sample plane to form a square lattice intensity pattern with a period $P$. (b)~Bird's-eye view of an array of nanoholes (intermediate regime) patterned from Si film with $h =` 70$~nm. SEM image, scale bar is equal to 1~$\upmu$m. (c)~Bird's-eye view of an array of nanoparticles (Mie resonators regime) patterned from Si film with $h = 70$~nm. SEM image, scale bar - 1~$\upmu$m. (d)~Mechanism of dewetting using the interference intensity pattern as a template with respect to the laser energy density. Si film with thickness $h$ reshapes via surface diffusion and hydrodynamic flow into a hemispherical shape with radius $R$ to minimize its surface energy. During heating, a-Si crystallizes into p-Si.}
\label{fig1}
\end{figure}

The laser interference was realized using a picosecond laser with a pulse duration of $\tau = 300$~ps and a wavelength $\lambda = 532$~nm, generated by a second-harmonic generation from a 1064~nm laser system, as shown in Figure~\ref{fig1}(a). The laser beam was split into four beams by a diffractive optical element. The four in-phase beams were transferred to the sample plane by a confocal optical system~\cite{voisiat2011flexible, maznev1998make}, ensuring equal incidence angles $\theta$ of each beam with respect to the surface normal of the sample (see Methods for more details on experimental setup). This results in a cos-shaped intensity pattern periodic in the $x$- and $y$-directions, see illustration in Figure~\ref{fig1}(a). Such intensity pattern is used to obtain nanostructures with corresponding spatial arrangement. The experimental demonstration is done using an amorphous Si films, deposited on glass substrates (see Supporting Information, S1, for dispersion parameters). By the increase of laser energy density $F$ with respect to the material threshold, different nanostructures can be obtained. In this work, we will focus on the regime of Mie-resonant nanoparticles (Mie resonators), see Figure~\ref{fig1}(c), considering nanohole-like nanostructures as the intermediate regime, see Figure~\ref{fig1}(b). 

Short-pulse lasers are known for a high-peak-power, which enables efficient micro- and nanostructuring of materials~\cite{wellershoff1999role}. However, several other considerations have to be made in the selection of the optimal source. The different physical phenomena involved in generating the nanopatterned surface will be explained in the following section. First, a thermal diffusion is playing a crucial role when the interference period goes down to the submicrometer range~\cite{gedvilas2018nanoscale}. The thermal diffusion length is proportional to the pulse duration~\cite{matthias1994influence}, thus a short pulse duration is required for small features to be obtainable. The pulse duration of $\tau=300$~ps was chosen to ensure a short laser-matter interaction time, but still produce a significantly large patterned area, as the shorter the pulses, the smaller the interference area over which they overlap~\cite{maznev1998make}. Second, the lattice constant (period) of the interference pattern $P$ is a wavelength-dependent parameter. The period $P$ of the four-beam interference pattern is defined by the wavelength of the laser radiation $\lambda$ and the incidence angle of the interfering beams $\theta$~\cite{kondo2003multiphoton}:
\begin{equation}
P=\frac{\lambda}{\sqrt{2}\sin{\theta}}~.
\end{equation} 
In the experimental demonstration, all four laser beams were incident at $\theta \approx 41^{\circ}$ with respect to the surface normal of the sample, thus a submicrometer period $P = 570$~nm was obtained, which is sufficient for the aforementioned applications in the visible and near infrared spectral range (see Supporting Information, S2, for more details on angle tunability). 

The intensity peaks of the interference pattern correspond to hotspots, as indicated in Figure~\ref{fig1}(d). Here, more energy is deposited in the Si film by absorption and it heats up faster than in the interference minima, leading to melting. As surface tension of the molten material is a temperature dependent parameter, it is lower in the interference maxima (hotspots) and is higher in the interference minima (colder zones)~\cite{millot2008surface}. Subsequently, Si flows from the interference maxima to the minima due to surface tension. If the material melting temperature is reached, the mass transport is also governed by a hydrodynamic flow as described by the mechanism of liquid-state dewetting~\cite{thompson2012solid, ye2019dewetting}. The irradiated material exhibits large temperature gradients and using the same interference pattern, but controlling the laser energy density with respect to the material damage threshold, we may obtain diverse patterns~\cite{voisiat2011flexible, indrisiunas2013two}. In case of thin films, the film-substrate interfacial free energy~$\gamma_{\text{fs}}$ determines the shape of the surface. The morphology of isotropic films is defined by Young’s equation~\cite{young1805, wyart1990drying}:
\begin{equation}
\cos\vartheta=\frac{(\gamma_{\text{s}}-\gamma_{\text{fs}})}{\gamma_{\text{f}}}~,
\end{equation} 
where $\vartheta$ is the equilibrium contact angle, $\gamma_\text{f}$ and $\gamma_\text{s}$ are the surface tensions of the film and the substrate, respectively. When heated to sufficiently high temperatures, the continuous film breaks into nanohole-like periodic structures, shown in Figure~\ref{fig1}(b). If the laser energy density is increased further, the film agglomerates to minimize the total free energy of the system~\cite{jiran1990capillary}, thus forming an array of hemispherical nanoparticles ($\vartheta=90^{\circ}$), see Figure~\ref{fig1}(c). Furthermore, if the ablation threshold is reached, a part of the sample is locally ablated and some of the material gets redeposited on the surface. The redeposited particles are spread irregularly based on the chaotic nature of evaporation and are significantly smaller compared to the ordered elements of the metasurface, thus their influence on the optical response is negligible. The ablation does not have a prominent influence unless the majority of the material is evaporated, subsequently destroying the ordered metasurface.

\begin{figure}[!t]
    \centering
    \includegraphics[width=82.5mm]{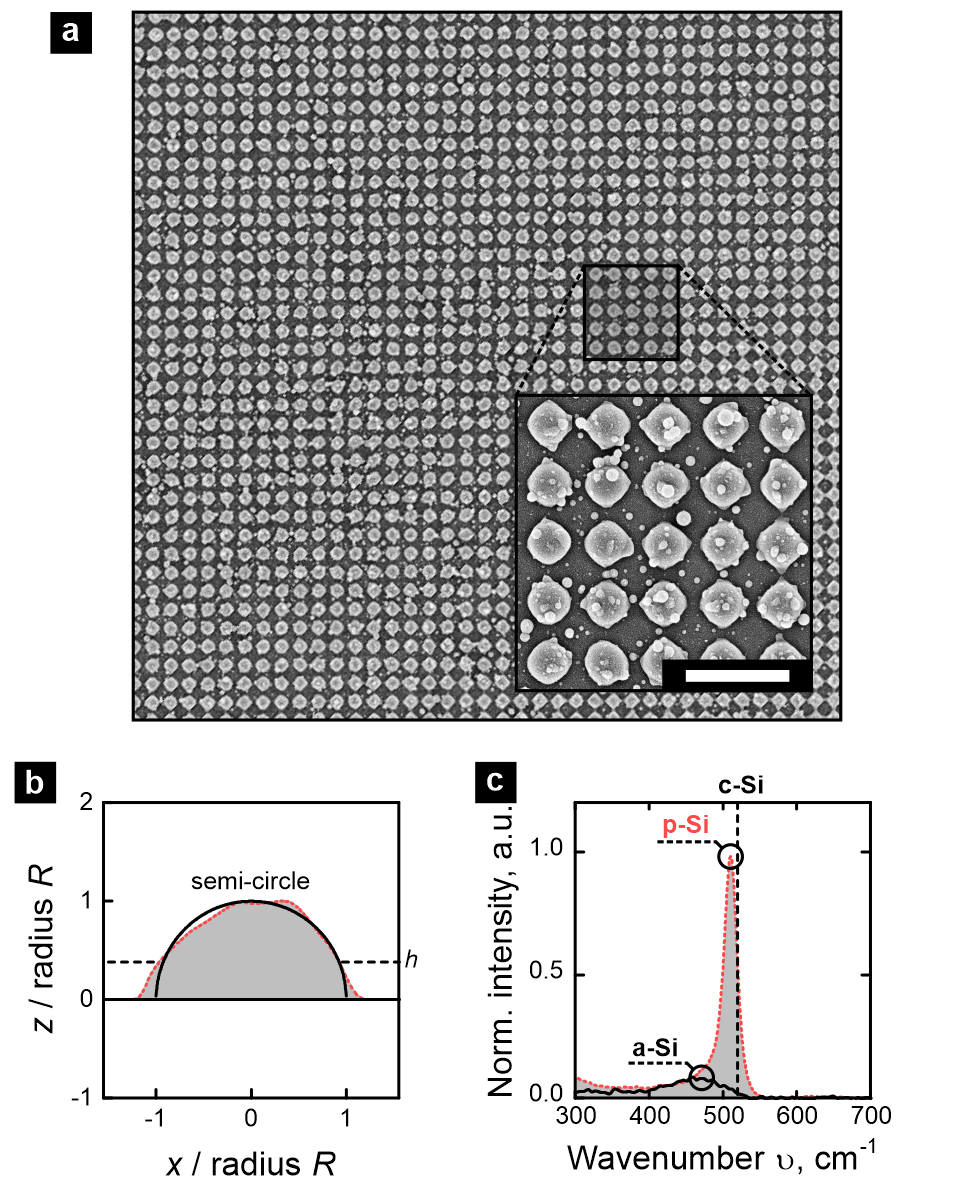}
    \caption{General properties of the fabricated metasurfaces. (a)~SEM image of a $20\times20$~$\upmu$m$^2$ area, constituted of Si-based Mie resonators. Scale bar is equal to 1~$\upmu$m. (b)~Cross-section of a single Mie resonator (red dotted line), compared to initial Si film with thickness $h = 70$~nm (black dashed line). The profile is fitted by a semi-circle (black solid line). (c)~Raman spectra indicating the transition from amorphous Si film with a peak at $\upsilon = 480~\text{cm}^{-1}$ (black line) to polycrystalline Mie resonators with a peak at $\upsilon = 510~\text{cm}^{-1}$ (red dotten line). The crystalline Si (c-Si) peak is at $\upsilon = 520~\text{cm}^{-1}$ (black dashed line).  
    }\label{fig2}
\end{figure}

Properly chosen irradiation conditions result in a large-area Mie-resonant metasurface, see exemplary scanning electron microscope (SEM) image in Figure~\ref{fig2}(a). The dependence of the metasurface parameters on the laser irradiation is discussed in-detail in the following chapters. Surface measurements using an atomic force microscope (AFM) confirm the prediction of a hemispherical shape of the individual nanoparticles, as shown in Figure~\ref{fig2}(b) by a typical cross-section of a patterned 70-nm-thick Si film patterned using laser energy density $F = 3.75 \text{J/cm}^2$. The horizontal cross-section of a single nanoparticle is approximated by a semi-circle.

In addition, the heat-affected amorphous Si tends to crystallize~\cite{becker2013polycrystalline,zograf2018local,iqbal1981raman,theodorakos2014picosecond}. The crystallization starts before the Si film reaches the melting temperature~\cite{zograf2018local}. Accordingly, the patterned area is polycrystalline, as confirmed by Raman spectroscopy measurements (note the Raman spectra of the structured and unstructured sample in Figure~\ref{fig2}(c)). For the unstructured Si film, a broad peak is observed at a Raman shift of $\upsilon = 480~\text{cm}^{-1}$, which corresponds to a Raman scattering of amorphous Si~\cite{theodorakos2014picosecond}. A peak at $\upsilon = 510~\text{cm}^{-1}$ appears only after the patterning and is attributed to the formation of nanocrystals~\cite{theodorakos2014picosecond}, while the peak of purely crystalline Si is expected at $\upsilon = 520~\text{cm}^{-1}$~\cite{iqbal1981raman,theodorakos2014picosecond}. Moreover, the Raman peaks become significantly sharper: from full-width at half-maximum (FWHM) of $\Delta\upsilon = 104.0~\text{cm}^{-1}$ in case of the unstructured film to $\Delta\upsilon = 23.6~\text{cm}^{-1}$ after exposure using a laser energy density of $F = 3.75~\text{J/cm}^2$.

\subsection{Patterning of large-area metasurfaces}

An essential advantage of direct laser interference patterning is the ability to obtain a relatively large area of periodic structures from a thin film or a bulk material by just a single laser pulse. In general, the envelope of the spatial energy distribution of the interference resembles the spatial energy distribution of the initial Gaussian beam~\cite{elkhoury2018utilizing}, while the interference area is limited by the spatial and temporal overlap of the interfering beams~\cite{maznev1998make}. Due to energy-dependent photothermal effects, the size of the patterned area also depends on the laser energy density with respect to the material threshold. The total patterned area may be distorted due to ellipticity of the initial beam or aberrations in the optical setup, thus, for simplicity, an average diameter $d$ is used to define the total patterned area. The diameter $d$ increases towards the diameter of the interference spot $d_{\text{spot}}$ with the increase of the laser energy density. The diameter of interference spot $d_{\text{spot}}$ is defined as the diameter of the spatial energy envelope at $1/e$ level (see Supporting Information, S3), with the increase of the laser energy density. The nanostructured elements inside the area are described by their diameter $D$ and height $H$, as well as period $P$ of the rectangular lattice.

\begin{figure}[!t]
    \centering
    \includegraphics[width=140mm]{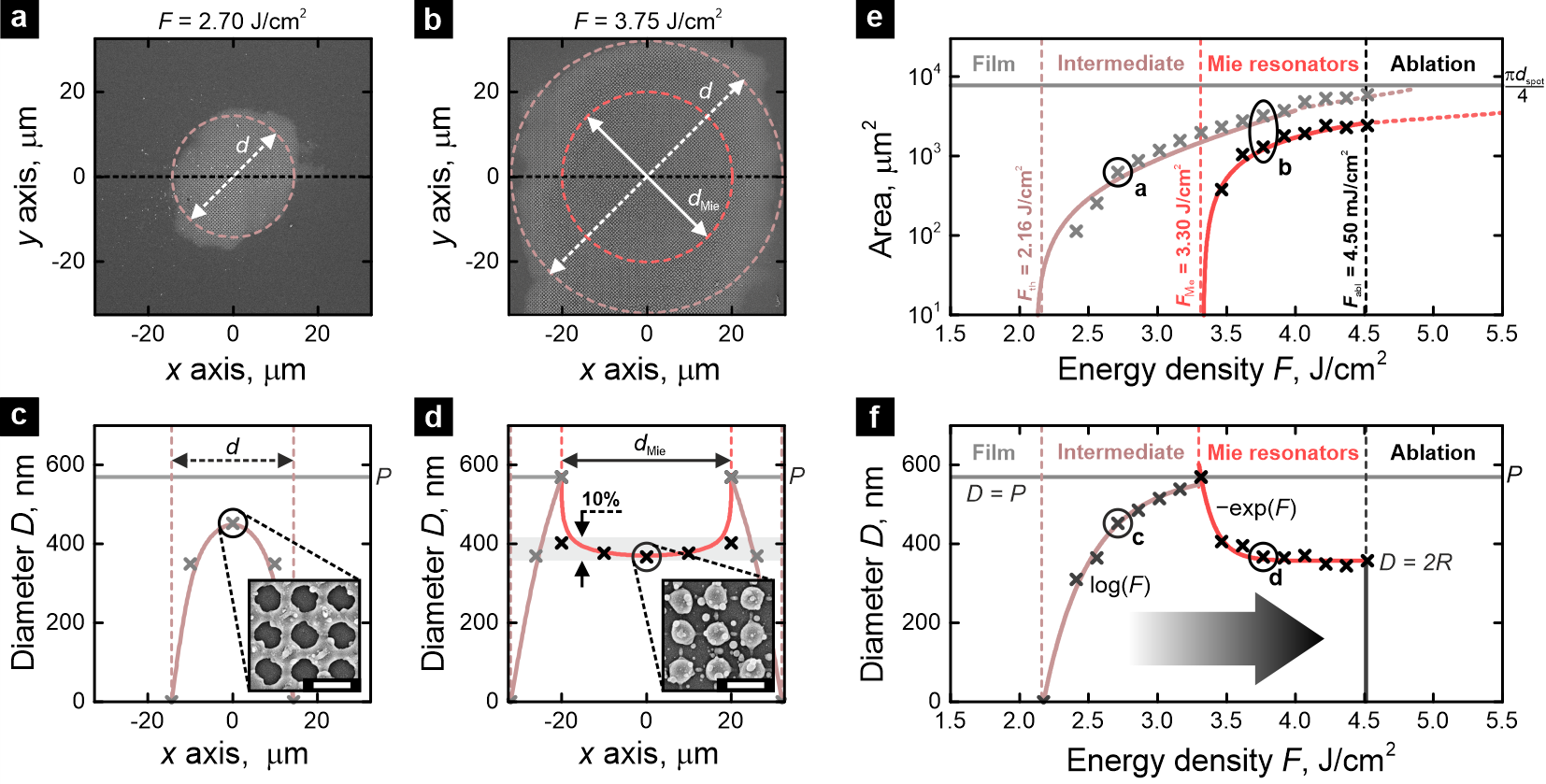}
    \caption{Large-area metasurfaces and their submicrometer elements from Si film with $h=70$~nm. (a)~A patterned area with diameter $d=28.6$~$\upmu$m (brown dashed line) using alaser energy density $F=2.70~\text{J/cm}^2$. (b)~A patterned area using a laser energy density $F=3.75~\text{J/cm}^2$. The diameter of the total area $d = 64$~$\upmu$m is highlighted by brown dashed line, while nanoparticle area $d_\text{Mie}=40$~$\upmu$m - red dashed line. (c)~Structural analysis of the area from (a), along $x$-axis. Nanoholes are formed with a maximum diameter $D=450$~nm, as shown in the SEM image, scale bar – 500~nm. (d)~Structural analysis of the area from (c), along $x$-axis. Size of nanoholes increases till nanoparticles are formed. Diameter of nanoparticles $D$ stays in $10~\%$ range off the central value, $D=2R=368$~nm. Inset shows SEM image, scale bar – 500~nm. (e)~Growth of the nanostructured area and the Mie-resonators area with respect to the applied laser energy density $F$. Thresholds are indicated for nanostructuring at $F_\text{th}=2.16~\text{J/cm}^2$, nanoparticles at $F_\text{Mie}=3.30~\text{J/cm}^2$, and ablation at $F_\text{abl}=4.50~\text{J/cm}^2$. (f)~Transition from Si film to Mie-resonators via analysis of the metasurface elements. Crosses indicate experimental data, brown line - logarithmic growth of nanoholes, red line - exponential decay of the nanoparticles size till $D=2R$.
    }\label{fig3}
\end{figure}

The first experimental demonstration is done using an amorphous Si film of $h = 70$~nm thickness. Si was deposited on a glass substrate (silicon dioxide, $n = 1.46$) by ion-beam deposition (see Supporting Information, S1, for dispersion parameters). In Figure~\ref{fig3}(a), we show a SEM image of a Si film irradiated by an arbitrarily selected laser energy density $F = 2.70~\text{J/cm}^2$. Using this energy, a part of the Si film reaches the material melting threshold, Si melts and is reshaped into an array of nanostructures, which can be seen as brighter color of the SEM image shown in Figure~\ref{fig3}(a). At slightly larger energy density $F = 3.75~\text{J/cm}^2$, we obtain a larger patterned area, see Figure~\ref{fig3}(b). 

As anticipated, the diameter of the total patterned area $d$ and the diameter of the area with Mie resonators $d_\text{Mie}$ is coupled to the applied laser energy density $F$, as shown in Figure~\ref{fig3}(e). It steadily grows towards the diameter of the interference spot $d_{\text{spot}} = 99$~$\upmu$m obtained from the pulsed beam spot size measurements~\cite{liu1982simple} (see Supporting Information, S3, for steps to estimate the interference spot and the thresholds). The threshold of the patterning process for a Si film with thickness $h = 70$~nm was found to be at the laser energy density $F_\text{th} = 2.16~\text{J/cm}^2$, while nanoparticles start to form above the laser energy density of $F_\text{Mie} = 3.30~\text{J/cm}^2$. As depicted in Figure~\ref{fig3}(e), for high laser energies up to a half of the patterned area consist of nanoparticles. Such area with $d_\text{Mie} = 55$~$\upmu$m consists of almost 7500 nanoparticles distributed in a rectangular lattice with a period $P = 570$~nm. Considering this is achieved using only a single laser pulse, significantly higher fabrication efficiency is demonstrated compared to the pulse-by-pulse direct laser writing (see head-to-head comparison in Supporting Information, S4).

The development of the metasurface elements was investigated by observation of the central part of the patterned area while changing the applied laser energy density $F$, see Figure~\ref{fig3}(f), where we plot the diameter $D$ of the nanostructures depending on the laser energy density $F$. We note a logarithmic growth of the size of the nanoholes starting from the threshold of $F_\text{th} = 2.16~\text{J/cm}^2$ up to the laser energy density of $F_\text{Mie} = 3.30~\text{J/cm}^2$. Increasing the laser energy density $F$, the photothermally induced heat is sufficient to melt the Si film within the whole range of the unit cell. At such condition, the surface of the Si film is discontinued and an array of round nanoparticles is formed due to the surface tension, as in the previously introduced model of templated solid-state dewetting~\cite{lian2006patterning, naffouti2016templated, fowlkes2011self, ye2011templated}. In case of nanoparticles (Mie resonators), the further increase of the laser energy density $F$ steadily decreases the diameter $D$ of the nanoparticles till $D = 2R$. The $h = 70$~nm Si film was converted into hemispherical nanoparticles with a diameter $D = 2R = 368\pm10$~nm (from $F \approx 3.45~\text{J/cm}^2$ to $F \approx 4.50~\text{J/cm}^2$), as obtained from analysis of SEM images and AFM measurements (see Methods for more details on sample characterization). At and beyond the critical laser energy density $F$, which was found to be at $F_\text{abl} \approx 4.50~\text{J/cm}^2$, the Si film is destroyed, as material is ablated at the center of the interference spot.

A challenge for direct laser interference patterning is the homogeneity over a large area. It is fundamentally limited by the Gaussian spatial energy distribution of the laser beam. Nevertheless, even without a complex beam shaping approach, a significant part of the patterned area is uniform. For example, the patterning with the laser energy density $F = 2.70~\text{J/cm}^2$ is at the fringe of the nanoparticle (Mie resonator) regime. The total patterned area with a diameter of $d = 28.6$~$\upmu$m, is constituted only of nanoholes. Their diameter variation at the central cut of the patterned area is indicated in Figure~\ref{fig3}(c). The diameter of the largest nanoholes in the center is equal to $D = 450\pm12$~nm, while it decreases fast when going away from the center till the nanoholes are no longer obtained due to insufficient laser energy. In contrast, the patterned film using a larger laser energy density $F = 3.75~\text{J/cm}^2$ exhibits the regime of Mie resonators, see Figure~\ref{fig3}(d). The Mie resonators seem to be spread more uniformly as all of the obtained structures in the area of $d_\text{Mie} = 40$~$\upmu$m fit into a $10~\%$ range off the value at the central part, from $D_\text{min} = 368\pm10$~nm to $D_\text{max} = 402\pm22$~nm.

It is worth to note that a metasurface of tens of micrometers in lateral size is already applicable in Raman spectroscopy, fluorescence or refractometry measurements, and, if required, has the potential to be spatially extended further by a partial interference spot overlap~\cite{indrisiunas2015direct,aguilar2018micro}.

\subsection{Tailoring of metasurface elements}

In the intent to access a large parameter set of Mie-resonant metasurfaces, the previously described methodology was repeated for amorphous Si films of different thickness $h$. In total, the Si film was deposited with the following thicknesses: 30~nm, 50~nm, 70~nm, and 90~nm. The patterning was carried out using the same experimental conditions (see Methods for more details on experimental setup). At each thickness $h$, the laser energy density~$F$ was gradually increased till the consecutive development of the Mie resonators. 

\begin{figure}[!t]
    \centering
    \includegraphics[width=105mm]{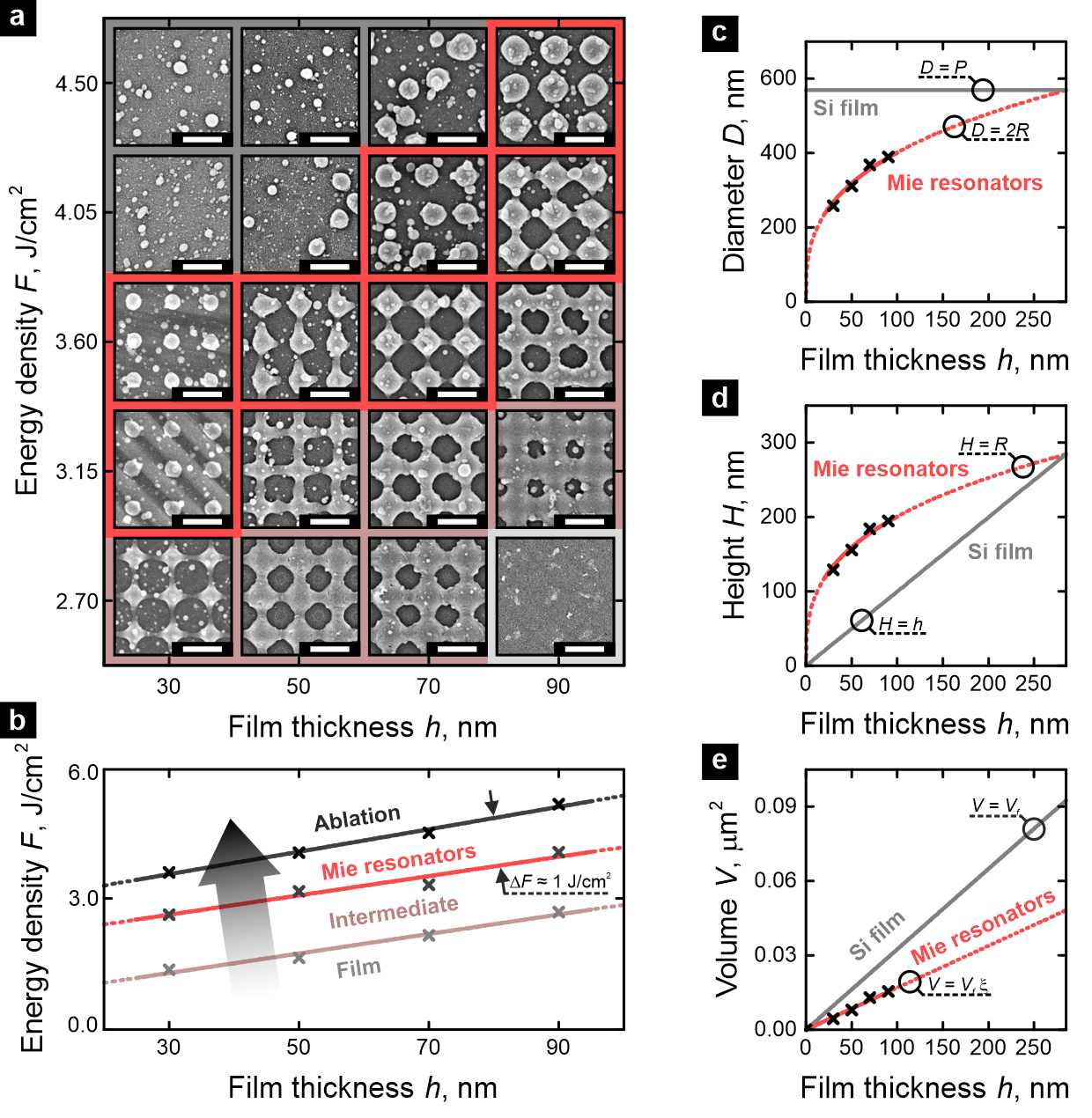}
    \caption{Tailoring of Si metasurfaces by film thickness and laser energy density. (a)~SEM images of patterned samples in a matrix of the energy density and the film thickness $h$. (b)~Laser energy density required for different regimes. Experimental points (crosses) are fitted linearly: brown line depicts the threshold of structuring, red line shows the start of the Mie resonators regime, which goes up to the ablation, see black line. (c)~Diameter of the fabricated Mie resonators $D$ as a function of film thickness $h$ (red line). The increase of energy density $F$ results in reshaping of the film (grey line) into a hemisphere shape. (d)~Height of the fabricated Mie resonators $H$ as a function of film thickness $h$: from a Si film with thickness $h$ (grey line) to a hemisphere with radius $R$ (red line). Cube-root function is extrapolated from experimental points (crosses) with $\xi = 52~\%$. (e)~Mie resonator volume relation to the initial Si film thickness (red line), which is a fraction $\xi = 52~\%$ of the volume of Si film within a unit cell (grey line).
    }\label{fig4}
\end{figure}

The process of the direct laser interference patterning is dependent on the film thickness~$h$, as is the laser ablation of thin films using a single laser beam~\cite{domke2014understanding, matthias1994influence}. Even-though the formation of metasurfaces was successful for all of the selected film thicknesses $h$, as shown in the SEM images of the samples patterned at explicitly chosen energy densities $F$, see Figure~\ref{fig4}(a), the experimental results differ for varying thickness $h$. For example, a laser energy density $F \approx 3.15~\text{J/cm}^2$ is sufficient for the formation of Mie resonators from a Si film with $h = 30$~nm, but such regime is not reached in case of thicker films. 

The experimentally obtained threshold for the intermediate regime, the regime of Mie resonators, and the ablation of Si for all of the thicknesses are indicated in Figure~\ref{fig4}(b), where the experimental points are fitted by a linear function of the film thickness. According to the well-known Beer-Bouguert-Lambert law, the intensity of electromagnetic wave decays exponentially while it propagates through a material. The penetration depth in Si at which the intensity of radiation decreases to $1/e$ of the incident intensity is $\delta \approx 5$~$\upmu$m, based on the absorption coefficient $\alpha = 2.0729\times10^5~\text{cm}^{-1}$ at radiation wavelength $\lambda = 532$~nm~\cite{pierce1972electronic}. As long as the thickness of the thin film is significantly smaller than the penetration depth, $h \ll \delta$, the assumption of a homogeneous energy distribution along the film thickness $h$ can be made~\cite{domke2014understanding}. A precise designation of the patterning regimes is done. As shown in Figure~\ref{fig4}(b), the range of the laser energy density for the Mie-resonant nanostructures to be obtained is $\Delta F \approx 1~\text{J/cm}^2$, which is the difference between the onset of the formation of the Mie resonators and the onset of ablation.

Next, an in-depth geometry analysis of the metasurfaces is performed. The high index metasurfaces were obtained with their elements varying in diameter $D$ and height $H$, depending on the film thickness $h$ and the laser energy density $F$. The geometry parameters after the patterning were determined by analysis of SEM images and complimentary AFM measurements (see Methods for more details). A typical measured cross-section of a unit cell is shown in Figure~\ref{fig2}(b). As the patterned Mie resonators are in the shape of a hemisphere, the cross-sections were approximated by a function of a semi-circle, with its radius $R = H = D/2$. This resulted in Mie resonators with their radius $R$ equal to $129\pm6$~nm, $156\pm8$~nm, $184\pm9$~nm, and $195\pm10$~nm, when patterned from the Si films with thickness $h$ equal to 30~nm, 50~nm, 70~nm, and 90~nm, respectively. 

As can be visually identified in Figure~\ref{fig4}(a), the patterned metasurface elements increase in size with the increase of the initial film thickness, which is also predicted by a direct volume transformation: 
\begin{equation}
V=V_\text{f}\cdot \xi~,
\end{equation}
where the volume of a single hemisphere $V$ is equal to a certain fraction $\xi$ of the film volume within the unit cell, $V_\text{f}=P^2\cdot h$. Subsequently, the radius $R$ of the Mie resonator can be defined by a function:
\begin{equation}
R=\sqrt[3]{\frac{3}{2\pi} \cdot P^2 \cdot h \cdot \xi}~,
\end{equation}
in our case, with the period $P$ being fixed and a cube-root relation to the film thickness $h$. The relation of diameter $D$ and height $H$ to the initial thickness $h$ of the Si film are shown in Figure~\ref{fig4}(c) and Figure~\ref{fig4}(d), respectively. In particular, the same trend shown in Figure~\ref{fig2}(f) is found for all thicknesses $h$. By increase of laser energy density $F$, the diameter of the Mie resonators $D$ changes from the size of the unit cell (period) $P$ to the radius of a hemisphere $R$, see Figure~\ref{fig4}(c), while the height $H$ changes from the initial film thickness $h$ to the radius $R$, see Figure~\ref{fig4}(d). Based on the cube-root function, it is determined that at least half of the Si film volume is transformed into the Mie resonators, $\xi \approx 52\%$, whereas the other part either gets fully evaporated or is redeposited on the surface. The predicted volume relation to the initial Si film thickness is plotted in Figure~\ref{fig4}(e).

Here, it should be noted that the volume fraction $\xi$ is not equivalent to the efficiency of the patterning, meaning the amount (mass) of the material transformed into Mie resonators. The density of the porous amorphous Si is $5$-$25~\%$ lower than the crystalline density~\cite{moss1969evidence}, thus the actual efficiency of Si film patterning into Mie resonators using single-pulse laser interference patterning with a pulse duration of $\tau = 300~$ps is $55$-$65~\%$. 

On a side note, in the necessity of having multiple metasurfaces with different-size elements on the same sample, one may want to control the radius $R$ of the Mie resonators by changing the period $P$, i.e. changing the radiation wavelength $\lambda$ or the incidence angle $\theta$. The wavelength change can be achieved by using achromatic optical elements and different laser harmonics or a tunable-wavelength source. For more tunability of the angle of incidence $\theta$ one may consider a different interference setup like the beam splitter and mirrors setup~\cite{oliveira2013sub}, though it is less robust in the spatial and temporal alignment (further discussion on period tunability is in Supporting Information, S2).

\subsection{Spectral analysis and Mie resonances}
\begin{figure}[!t]
\centering
\includegraphics[width=82.5mm]{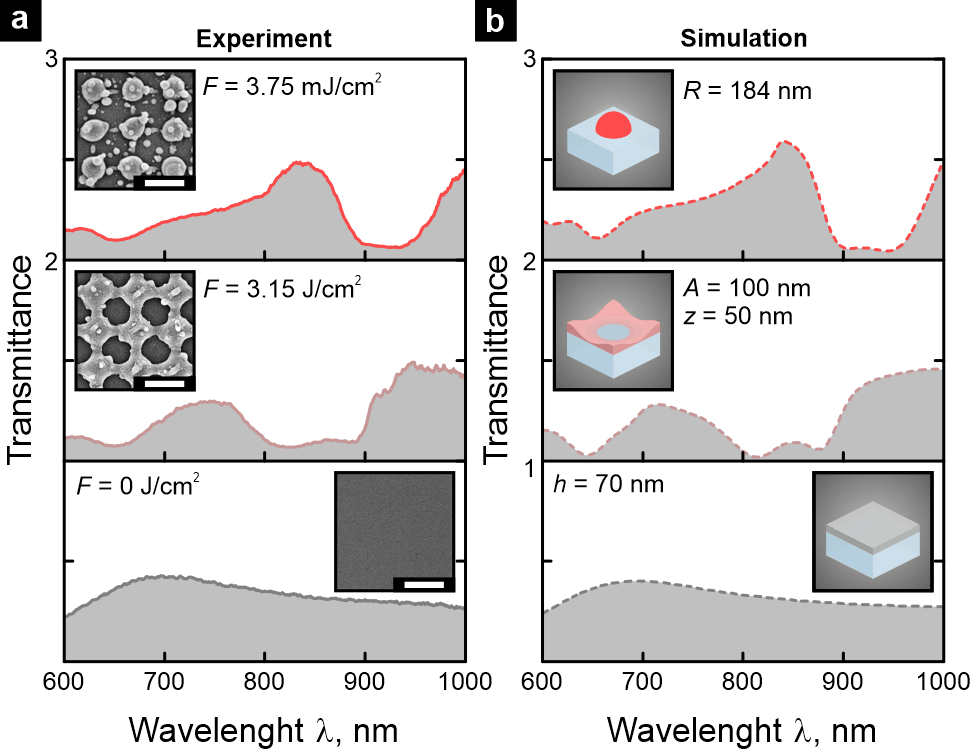}
\caption{Spectral response of Si film prior and post patterning. (a)~Measured spectra of unstructured $70$~nm-thick film, nanoholes (intermediate regime) and nanoparticles (Mie resonators), obtained by irradiation of $F = 3.15~\text{J/cm}^2$ and of $F = 3.75~\text{J/cm}^2$, respectively. Insets show SEM images from the central parts of the patterned areas, scale bars are equal to 500~nm. (b)~Numerically obtained spectra of the corresponding structures: film ($h = 70$~nm), nanoholes (sinusoidal surface with amplitude $A = 100$~nm, positioned $z = 50$~nm above the substrate), and hemispherical Mie resonators ($R = 184$~nm). The insets show sketches of the unit cells.}\label{fig5}
\end{figure}

The optical response of Si film is wavelength-dependent, but even sharper spectral features appear after the patterning. This is demonstrated by the measured transmittance prior- and post-patterning of a Si film with thickness $h = 70$~nm, see Figure~\ref{fig5}(a). Here, the transmittance of the film is compared to the transmittance of the samples irradiated by laser energy densities $F = 3.15~\text{J/cm}^2$ and $F = 3.75~\text{J/cm}^2$, which correspond to the intermediate regime and the Mie resonator regime, respectively. 
In Figure~\ref{fig5}(b), the measured spectra are supported by the full-wave simulations. The nanoparticles (Mie resonators) are defined as hemispheres with a certain radius $R$, while the nanoholes are defined by a two dimensional sinusoidal surface with a period $P$, a mean distance from the substrate $z$, and an amplitude of the sinusoidal wave $A$ (see Methods for more details). The refractive index of Si is fitted between the refractive indices of amorphous Si (a-Si) and crystalline Si (c-Si) due to its dependence on the crystallization state (see Supporting Information, S1, for dispersion parameters).

Due to the sinusoidal interference intensity distribution, the sidewalls of the nanoholes are inclined and uneven across the surface, see the sketch in Figure~\ref{fig5}(b). Accordingly, it is difficult to assign the peaks or the dips of the transmission spectra in Figure~\ref{fig5}(a,b) to a particular type of resonance, as they could be of a mixed nature, including but not limited to guided-mode resonances~\cite{ko2018wideband}, and Mie resonances~\cite{groep2013designing, staude2013tailoring}, and further investigation beyond the scope of this paper is required. In contrast, in case of the well-defined-shape Mie resonators the dips in the transmission spectra, such as a clearly pronounced dip at $\lambda \approx 950$~nm, can be assigned to Mie-type electric and magnetic resonances, as known from our previous research~\cite{berzins2019submicrometer}.

\begin{figure}[!t]
    \centering
    \includegraphics[width=82.5mm]{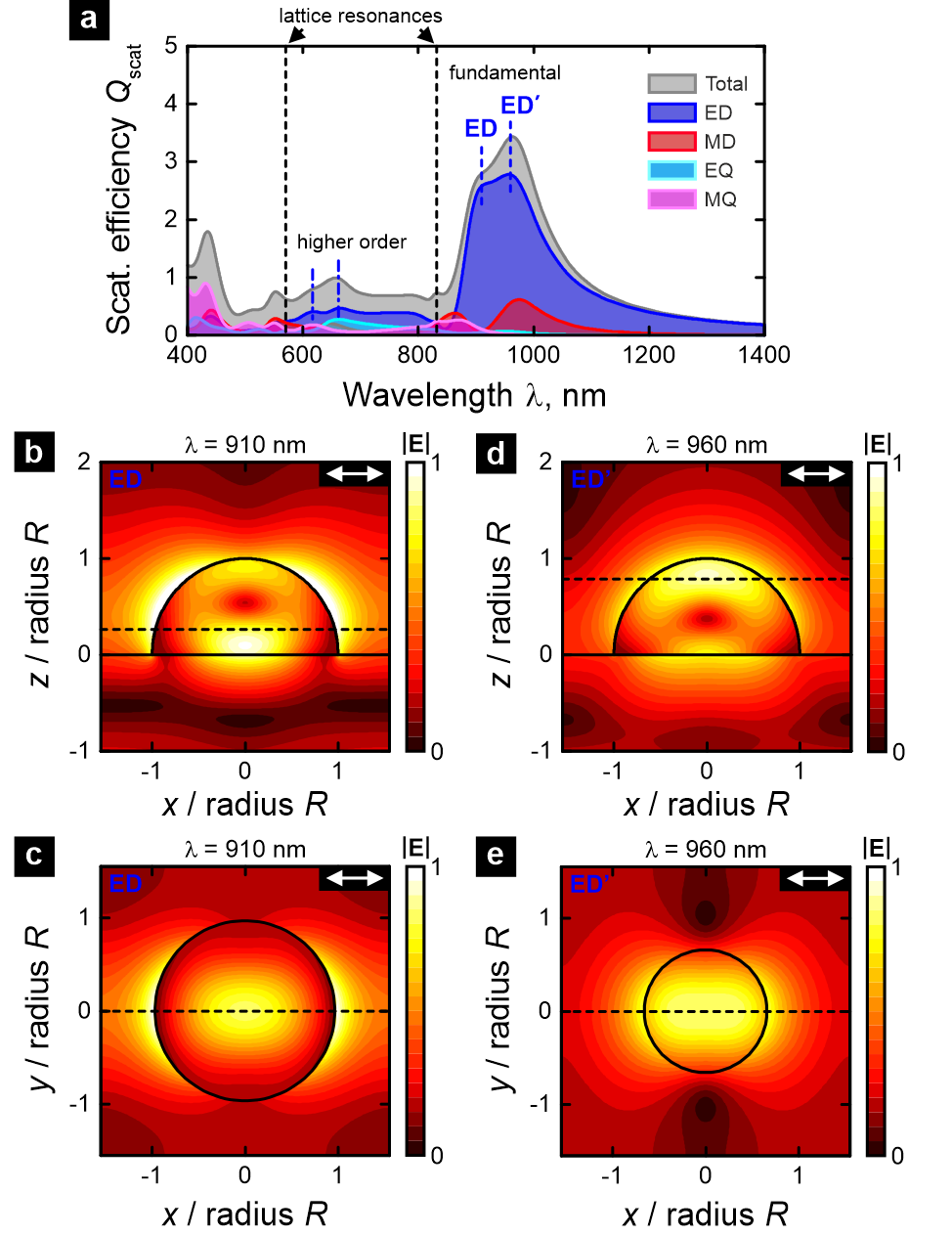}
    \caption{Mode decomposition. (a)~Scattering efficiency $Q_\text{scat}$ of metasurface consisting of Mie resonators with $R = 184$~nm (patterned from Si thin film of $h = 70$~nm): total scattering and contributions from electric dipole (ED), magnetic dipole (MD), electric quadrupole (EQ), and magnetic quadrupole (MQ) modes. The ED resonance is split into ED and ED’ because of symmetry breaking. Lattice resonances are indicated by black dashed lines. (b)~Electric field at a vertical cross-section of a unit cell at $\lambda = 910$~nm, corresponding to ED. (c)~Electric field at a vertical cross-section at $\lambda = 960$~nm, corresponding to ED’. (d)~Electric field at a horizontal cross-section at $\lambda = 910$~nm, corresponding to ED. The cut is at 1/4 of the height $H$, as shown in (b). (e)~Electric field at a horizontal cross-section at $\lambda = 960$~nm, corresponding to ED’. The cut is at 3/4 of the height $H$, as shown in (c). Electric field is $x$-polarized, its magnitude is normalized to 1. 
    }\label{fig6}
\end{figure}

For the Mie-type resonators, the origin of the strong scattering at $\lambda \approx 950$~nm is assigned to electric dipole (ED) based on the mode decomposition shown in Figure~\ref{fig6}(a). More precisely, the ED resonance is split into two resonances due to symmetry breaking when using optically non-identical materials for substrate and cladding, e.g. using a glass substrate in air~\cite{butakov2016designing}. Even-though these the resonances are partially overlapping, the distinction between them can be seen by observation of the electromagnetic fields inside the nanostructures shown in Figure~\ref{fig6}(b-e), where the electric field distributions in the horizontal and vertical cross-sections of the Mie resonator are given. One of the ED resonances is identified at $\lambda = 910$~nm with the strongest field being at the bottom of the hemisphere, see Figure~\ref{fig6}(b,c), while another peak, marked as ED’, lies at $\lambda = 960$~nm with electric field concentrating at the top of the structure, see Figure~\ref{fig6}(d,e).

The ED resonances originate from the collective polarization induced in the dielectric nanoparticle, while magnetic dipole (MD) resonances are driven by the electric field coupling to the displacement current loops. However, the latter are not as prominent when compared by the scattering efficiency $Q_\text{scat}$ (scattering cross-section normalized to geometric cross-section), because of a relatively short height $H=D/2$ and intrinsic Si losses. Moreover, a suppression of the ED scattering at $\lambda \approx 850$~nm may relate to a non-scattering anapole mode~\cite{miroshnichenko2015nonradiating, gurvitz2019}. The optical response at shorter wavelengths is influenced by electric quadrupole (EQ) and magnetic quadrupole (MQ), see Figure~\ref{fig6}(a). Lattice resonances are indicated at the wavelengths of $\lambda = 570$~nm and $\lambda = 832~$nm, based on the period $P$ and the refractive index of air ($n = 1$) and glass ($n = 1.46$), respectively. Octupoles and other higher order modes are neglected.

\begin{figure}[!t]
    \centering
    \includegraphics[width=140mm]{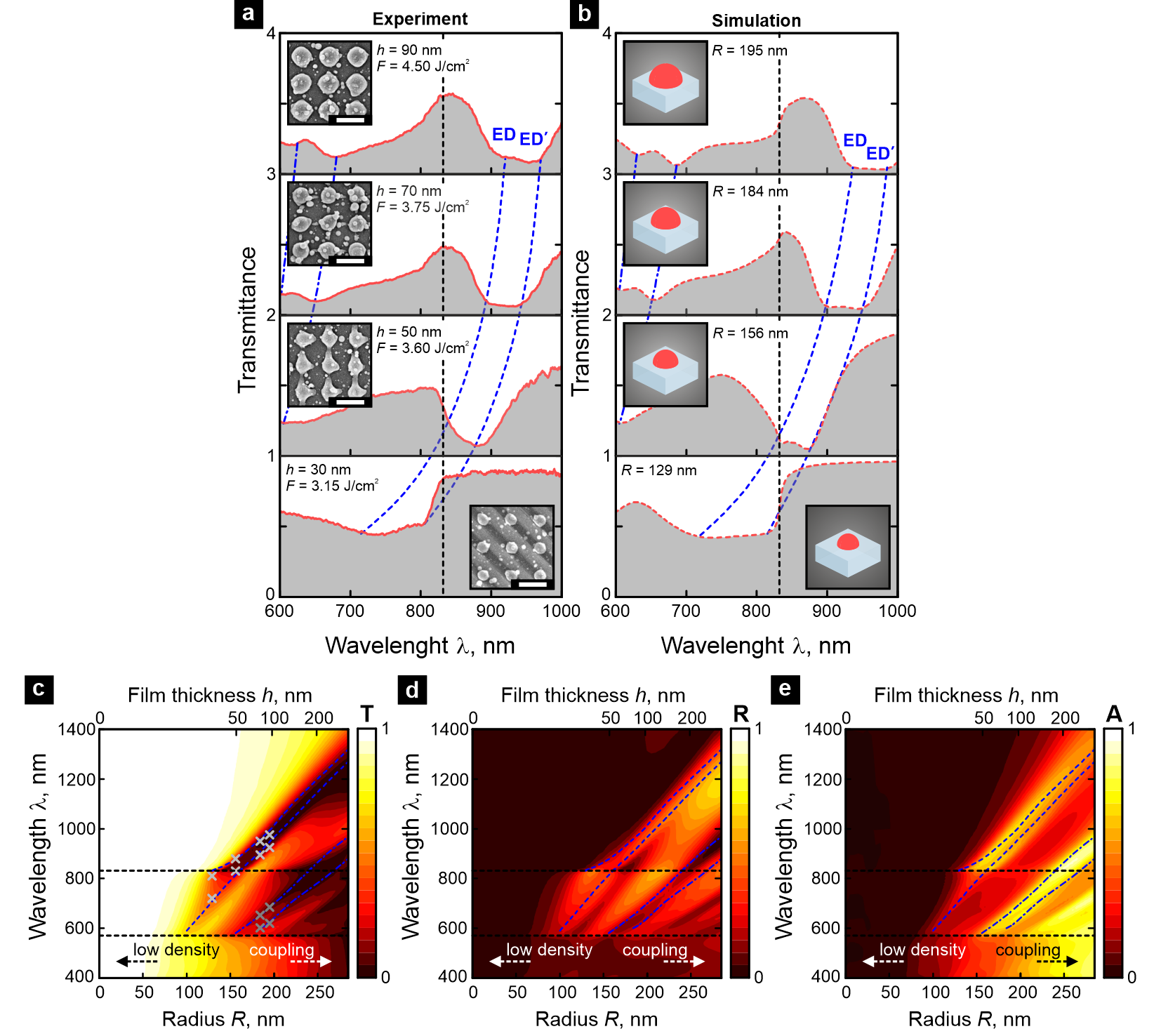}
    \caption{Spectral response of the Mie-resonant metasurfaces depending on their size. (a)~Transmittance of the samples fabricated from Si films with varied thickness: 30~nm, 50~nm, 70~nm, and 90~nm. SEM images are shown in insets, scale bars depicts 500~nm. Dominant Fundamental electric dipole modes, ED and ED’, are highlighted by blue dashed lines, higher order – by blue dash-dot lines. (b)~Numerically obtained spectra of the Mie resonators. Electric dipole modes are highlighted. Insets show sketches of resonators with a certain radius $R$. (c)~Extrapolated transmittance map in dependence of Mie resonators’ radius $R$ and the initial film thickness $h$. Electric dipole resonances are highlighted, crosses depict experimentally obtained transmittance dips. Similarly simulated maps of: (d)~reflectance, (e)~absorbance. At small radius $R$ we have insufficient density of the resonators, while at high values – coupling between neighboring resonators. Black dashed lines shows lattice resonances. 
    }\label{fig7}
\end{figure}

The spectral positions of the resonances are related to the optical size of the nanostructures~\cite{groep2013designing}. In particular, a red-shift of the transmittance dips is observed in the measured spectra with the increase of the initial film thickness and, subsequently, the size of the Mie resonators, see Figure~\ref{fig7}(a). The nanostructures patterned from Si film thicknesses of 30~nm, 50~nm, 70~nm, and 90~nm, and their respective spectral functions were successfully reproduced numerically using the measured geometry, see Figure~\ref{fig7}(b). The mode decomposition of the periodic resonators confirms the trend of the spectral functions and their correlation with the Mie resonances, note the highlighted fundamental electric dipole resonances ED and ED’ in both experimental and numerical results in Figure~\ref{fig7}(a,b).

As the measured transmittance and the numerical simulations have a strong agreement, this enables a reliable emulation of the spectral behavior (transmittance, reflectance and absorbance) within and even outside of the experimentally observed spectral range, plotted in Figures~\ref{fig7}(c-e). Here it can be noted that the influence of the resonances becomes apparent only when the radius of the Mie resonators is sufficiently large, $R > 100$~nm. Up to that point the structures fill only a very small portion of the surface, thus their density is too low considering the lattice configuration. In contrast, when $R > 250$~nm, we have another effect, because the gap between the neighboring metasurface elements reduces significantly, and causes strong coupling and subsequent broadening of the resonances. As a result, in the demonstrated case, the optimal range of the radius is from $R = 100$~nm to $R = 250$~nm, in which we observe a strong and wavelength-selective optical response. 

In perspective, the presented technique could be extended towards different and more complex shapes of the metasurface elements and their geometrical distribution by manipulation of phase~\cite{indrisiunas2013two, fernandez1998effects}, control of polarization~\cite{indrisiunas2017new}, or sub-period positioning of the sample~\cite{indrisiunas2015direct}. Also, as already indicated, the patterned area could be increased from tens of $\upmu$m$^2$ to cm$^2$, or even m$^2$, by a partial spatial overlap of the interference spots~\cite{indrisiunas2015direct, aguilar2018micro} with a stitching error as low as $9~\%$, which could also be reduced by the spatial modulation of the beams. High repetition rates of pulsed lasers and speeds of high-accuracy positioning stages already allow processing speeds within the mass-production scale, up to 1~m$^2$/min~\cite{lasagni2017direct}, while the fabricated large-scale uniform metasurface could be equipped with local functionalities by the complimentary process of photothermal reshaping~\cite{zhu2017resonant, berzins2020}.

Furthermore, besides the optical properties, the demonstrated nanostructures may serve several different roles based on their physical size and geometry. An array of Si nanoholes could be used as a physical matrix for test-subject deposition for surface enhanced Raman spectroscopy (SERS)~\cite{vzukovskaja2019rapid} or thousands of nanotubes for other sensing applications~\cite{escobedo2013chip, blanchard2017sensing}. Moreover, Si nanostructures may be used for mechanical bactericidal effect~\cite{ivanova2013bactericidal}, and other applications.

\section{Conclusions}

We have introduced a single-step technique for the fabrication of Mie-resonant metasurfaces. For the first time, direct laser interference patterning has been applied for fabrication of high-index nanostructure arrays on a glass substrate.

The area of the obtained metasurface spans across $10^2\text{-}10^4$~$\upmu\text{m}^2$ and, subsequently, consists of thousands of polycrystalline Si nanostructures distributed in an ordered rectangular lattice with a period $P = 570$~nm. The technique is carried out using only a single 300~ps laser pulse, thus shows a very large throughput compared to other existing laser-assisted fabrication techniques. Moreover, it was demonstrated that the hemispherical Mie resonators can be tailored by the pre-selection of the initial Si film thickness $h$ and control of the irradiation conditions. The variation of the film thickness allowed us to obtain Mie resonators with diameters $D$ ranging from less than 260~nm to almost 400~nm, with their electric and magnetic resonances covering the visible and the near infrared spectral range. 

Last but not least, the technique is transferable to other high-index dielectrics, e.g. titanium dioxide (TiO$_2$) and gallium phosphide (GaP). The presented analysis serves as a strong building block in bridging the gap between high-index dielectric metasurfaces and their implementation in mass-produced devices. 

\section{Methods}
\subsection{Experimental setup}
The experiments were done using a high-power picosecond laser (Atlantic HE, Ekspla, Ltd.) with a pulse duration of $\tau = 300$~ps and a repetition rate of $\nu = 1$~kHz. The fundamental wavelength $\lambda = 1064$~nm was transformed into its second harmonic $\lambda = 532$~nm to be in the high-absorption range of amorphous Si. The diameter of the Gaussian beam at the output was $d_\text{out} \approx 1.9$~mm at $1/e^2$, as measured with a CCD camera, but expanded by a factor of 2 using an optical telescope. The laser beam was divided into multiple beams using a multi-spot diffractive optical element (DOE) with a $10^{\circ}$  full angle (HOLOEYE Photonics AG). The 1st order beams were directed to the sample plane by using a system composed of two plano-convex spherical lenses ($L_1 = 100$~mm and $L_2 = 15$~mm), while the 0th diffraction order and the higher order beams were blocked by a metal aperture. The laser power was controlled by an attenuator based on a polarizer, a beam splitter and a heat sink. The setup was used to pattern Si films of four different thicknesses: 30~nm, 50~nm, 70~nm, and 90~nm, which were fabricated on top of glass substrates by ion-beam deposition. The samples were positioned using a translation stage (Aerotech, Inc.). All of the experiments were carried out in ambient conditions.

\subsection{Sample characterization}
The spectral analysis was done with a plane-wave illumination using an inverted optical microscope system (Axio Observer D1, Carl Zeiss AG) with an integrated broad-band VIS/IR imaging spectrometer (iHR320, HORIBA Jobin Yvon GmbH). In addition, SEM images were taken (Helios NanoLab G3 UC, FEI Co.) and the surface topology was measured by an AFM with incorporated peak force tapping technology (Dimension Edge, Bruker Co.). The Raman spectra were taken using a commercially available confocal Raman system (WITec GmbH) equipped with a 785~nm laser. The light was focused onto the sample and the scattered light was collected with the same microscope objective ($\text{NA} = 0.95$). The measurements were taken with laser power of 1~mW and integration time of 1~s with 5 accumulations. Measured spectra were background corrected using the statistics-sensitive non-linear iterative peak-clipping (SNIP) algorithm with 100 iterations. 

\subsection{Numerical simulations}
The transmittance was calculated using a finite-difference time-domain (FDTD) method (Lumerical, Inc.). A 3D model of an infinite array of amorphous Si nanostructures was emulated by a unit cell with periodic boundary conditions (PBC) on the sides and perfectly matched layers (PML) on the top and the bottom of the simulation domain. The unit cell was illuminated by a normally-incident linearly-polarized plane-wave source. Frequency domain field and power monitors were placed to record transmittance and reflectance. The Mie resonators were constructed as ideal hemispheres with a certain radius $R$. The intermediate regime of nanoholes was implemented by a sinusoidal layer, characterized by a period $P$ in the transverse directions, a distance from the glass surface $z$, and an amplitude $A$ of the sinusoidal wave. The refractive index of Si is given in Supporting Information, S1. The refractive index of glass was set to $n = 1.46$. The electromagnetic fields for the mode decomposition was calculated by a finite element method (COMSOL, Inc.). The scattering efficiency of various modes were computed from the current density (see Supporting Information, S5, for details). 

\section{Authors contribution}
J.B. developed the idea of using single-pulse laser interference for fabrication of Mie resonators. S.I. built the experimental setup. S.F. deposited the initial Si films. J.B. and S.I. carried out the experiments. M.S. took the SEM images and AFM measurements. J.B. carried out the linear spectroscopy measurement. J.B. performed the numerical simulations. K.V.E., A.N. and G.G. provided the mode decomposition. P.G., T.P., S.M.B.B., and F.S. participated in the discussions and supervision. J.B. wrote the manuscript. All authors analyzed the data, read and corrected the manuscript before submission.

\section{Acknowledgements}
The authors acknowledge financial support from the European Union’s Horizon 2020 research and innovation programme under the Marie Sklodowska-Curie grant agreement No.~675745 and German Federal Ministry of Education and Research (FKZ~03ZZ0434, FKZ~03Z1H534, FKZ~03ZZ0451). The authors express gratitude to O.~\v{Z}ukovskaja and D.~Cialla-May from Leibniz Institute of Photonic Technology Jena for the Raman spectroscopy measurements. The authors also thank B.~Voisiat, M.~Gedvilas and G.~Ra\v{c}iukaitis for valuable contribution developing the optical system and helpful discussions.

\section{Supporting Information Available}
The supporting information is not available in this version of the manuscript.

\section{Disclosures}
The authors declare no conflicts of interest.

\bibliography{refs}


\end{document}